\documentclass[a4paper,10pt,twoside]{levelmeter}
\usepackage{graphicx}
\usepackage{ctable}
\usepackage{booktabs}
\usepackage{amssymb,bm,mathrsfs,bbm,amscd}
\usepackage[tbtags]{amsmath}
\usepackage{lastpage}
\usepackage{CJK}
\begin{document}
\begin{CJK*}{GBK}{song}

\title{Development of the Liquid Level Meters for the PandaX Dark Matter Detector}
\pagestyle{plain}
\thispagestyle{fancy}          
\rhead{Submitted to `Chinese Physics C'}   

\author{
J. Hu$^{1}$, H. Gong$^{1}$, Q. Lin$^{1}$, K. Ni$^{1}$, A. Tan$^{2}$, Y. Wei$^{1}$, M. Xiao$^{1}$, X. Xiao$^{1}$, and L. Zhao$^{1}$}

\maketitle
\address{
$^1$ Department of Physics and Astronomy, Shanghai Jiao Tong University, Shanghai 200240, P.R. China\\
$^2$ Department of Physics, University of Maryland, College Park, MD 20742, U.S.A.
}
\begin{abstract}
The two-phase xenon detector is at the frontier of dark matter direct search. This kind of
detector uses liquid xenon as the sensitive target and is operated in two-phase (liquid/gas) mode, where the liquid
level needs to be monitored and controlled in sub-millimeter precision. In this paper, we present a detailed design
and study of two kinds of level meters for the PandaX dark matter detector. The long level meter is used to monitor
the overall liquid level while short level meters are used to monitor the inclination of the detector. These level meters are cylindrical capacitors custom-made from two concentric metal tubes. Their capacitance values are read out by a
universal transducer interface chip and recorded by the PandaX slow control system. We present the developments
that lead to level meters with long-term stability and sub-millimeter precision. Fluctuations (standard deviations) of less than 0.02 mm for the short level meters and less than 0.2 mm for the long level meter were achieved during a few days of test operation.
\end{abstract}



\section{Introduction}
Liquid xenon (LXe) nowadays is used as a suitable target for direct detection of the weakly interacting massive particles (WIMPs) dark matter \cite{lab1,lab2}. In a xenon detector operated in two-phase mode, the LXe level in the time projection chamber (TPC) is an important parameter for the experiment and the liquid level needs to be monitored and controlled in sub-millimeter precision \cite{lab3}. In order to monitor the liquid level precisely for the detector operation, level meters working at the LXe temperature (around 178~K) are constructed and the measurement methods are developed for the PandaX dark matter detector. These level meters are all custom-made concentric cylindrical capacitors.

\begin{center}
\includegraphics[width=7cm]{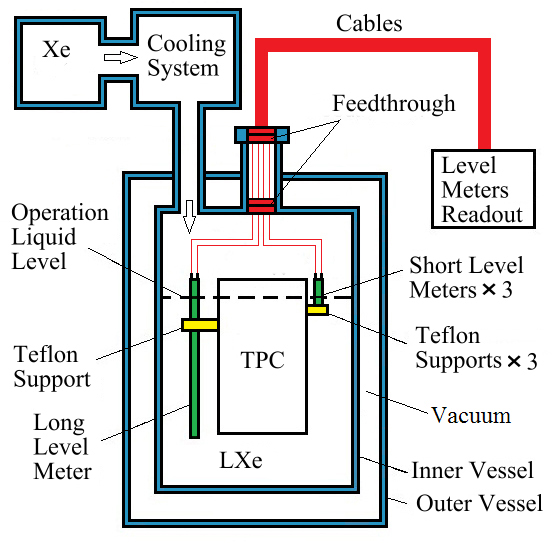}
\figcaption{\label{fig1}Schematics of the two-phase xenon detector as used in PandaX. LXe is contained in an inner vessel insulated by vacuum from the outside. One long liquid level meter monitors the overall liquid xenon height and three short level meters monitor the height of the liquid-gas interface around the TPC.}
\end{center}

As shown in Fig.~\ref{fig1}, a long level meter is used to monitor the total LXe level in the inner vessel of the detector, which is used during the filling and recovery of the liquid xenon. Three short level meters are mounted at the height of liquid-gas interface around the TPC to monitor its declination, which is important to maintain a uniform gas gap across the horizontal plane for homogeneous signals.  In this paper, we present the detailed design and method of the level meter fabrication and readout in section 2. The results during a few days of operation in a test xenon detector are reported in section 3. We conclude, in section 4, that the level meters can reach sub-millimeter precision and stability which are required by the detector operation.

\section{Design and Methods}
\subsection{Design of the level meters}
The short level meter and the long level meter are custom-made concentric cylindrical capacitors, made by two stainless steel tubes nested together with nylon fishing lines wound around the inner tube to separate it from the outer tube. In this study, the Universal Transducer Interface (UTI) sensor chip from Smartec Company \cite{lab4} are used as measurement device for the level meters. Considering the measurement range of the UTI sensors, the capacitances of level meters are limited to 12 pF for short level meters and 300 pF for the long level meter in LXe while the capacitance per unit length should be large enough to obtain a precise measurement. In addition, the dimensions of the level meters should be as small as possible to minimize their influence to the detector.

With these requirements, we design the liquid level meters using two concentric stainless steel tubes with an inner diameter of 4.5-mm for the outer tube and an outer diameter of 4.0-mm for the inner tube. The two tubes are separated by a 0.2-mm diameter nylon fishing line surrounded on the inner tube. The length of the short level meters are 12-mm, giving a capacitance of 10.61 pF when it's totally immersed in liquid xenon. The length of the long level meter is 254-mm, giving a capacitance of 232.3 pF when it's totally immersed in liquid xenon. The capacitance per unit length is 0.41 pF/mm for the short level meters and 0.44 pF/mm for the long one. Table~\ref{tab1} shows the dimensions of level meters and their theoretical capacitances when totally immersed in gas xenon ($C_{GXe}$) or liquid xenon ($C_{LXe}$). The effect of the nylon lines, which are uniformly distributed in the level meters and occupy 1.57~mm$^3$ in the short level meters and 9.42~mm$^3$ in the long level meter, is considered for the estimation of capacitances. The assembled level meters are shown in Fig.~\ref{fig2}.

\begin{center}
\tabcaption{ \label{tab1}Design parameters of the level meters. $d_1$ is the inner diameter of the outer tube. $d_2$ is the outer diameter of the inner tube. $L$ is the length. $d_f$ is the diameter of the nylon fishing line separating the two tubes. $C_{GXe}$ and $C_{LXe}$ are the theoretical capacitances assuming dielectric constant of 1.0 for gas xenon and 1.96 for liquid xenon~\cite{lab5} respectively. }
\footnotesize
\begin{tabular*}{150mm}{@{\extracolsep{\fill}}ccccccc}
\toprule level meters& ${d_1}$(mm) & ${d_2}$(mm)& $L$(mm) & ${d_f}$(mm) & ${C_{GXe}}$(pF)&${C_{LXe}}(pF)$ \\
\hline
$SLM$  &4.5 & 4.0 & 12  & 0.2 & 5.66& 10.61 \\
$LLM$ &4.5 & 4.0 & 254 &0.2 & 119.9 & 232.3 \\
\bottomrule
\end{tabular*}
\end{center}

\begin{center}
\includegraphics[width=8cm]{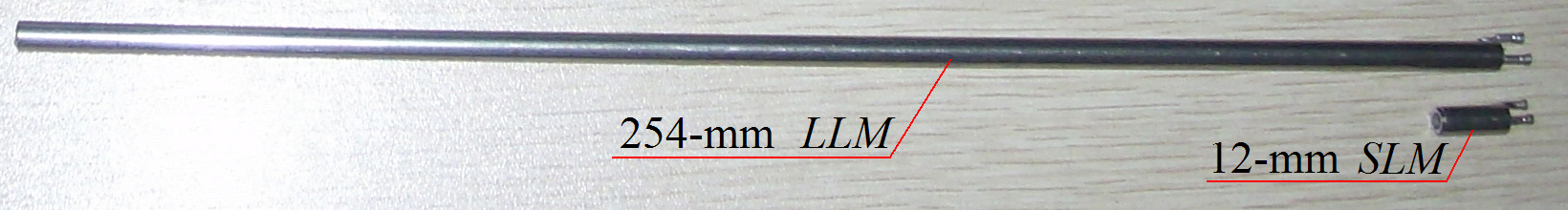}
\figcaption{\label{fig2}A photo showing an assembled 12-mm short level meter ($SLM$) and a 254-mm long level meter ($LLM$), made with two concentric stainless steel tubes.}
\end{center}

\subsection{Cable connections}
Since the relative large parasitic capacitance caused by long cables connected parallel to the capacitors, it's difficult to obtain a small capacitance accurately by direct measurement. The UTI sensor chip uses the principle of four-wire measurement to overcome the problem of parasite capacitances and three-signal technique for auto-calibration. The company provides an evaluation board which has several different measurement modes for capacitances, Pt-100 temperature sensors, electric bridges etc..\cite{lab4}. In this study, we use mode 2 to measure the three capacitors in the range of 0$\sim$12 pF for the three short level meters simultaneously and mode 4 to measure the single capacitor in the range of 0$\sim$300 pF for the long level meter.

The schematics to use mode 2 to measure the three capacitances for the short level meters are shown in Fig.~\ref{fig3}. A reference capacitor with unchanging capacitance is needed for this mode. We used a 12-pF reference capacitor made with the same method as the other short level meters but twice longer. It is placed in the gas xenon just above the liquid surface in order to cancel the interference in the readout cables and also to ensure no capacitance changes with the liquid level. The variation of the reference capacitor with temperature was measured from room temperature to liquid nitrogen temperature. The capacitance of the reference capacitor changes from 12.2 pF to 11.8 pF accordingly. The cables which connect the level meters to the UTI board go through the inner vessel and the outer vessel by vacuum feedthroughs and finally they are connected to the outside readout box with the Sub-Miniature-A (SMA) connectors.

\begin{center}
\includegraphics[width=15cm]{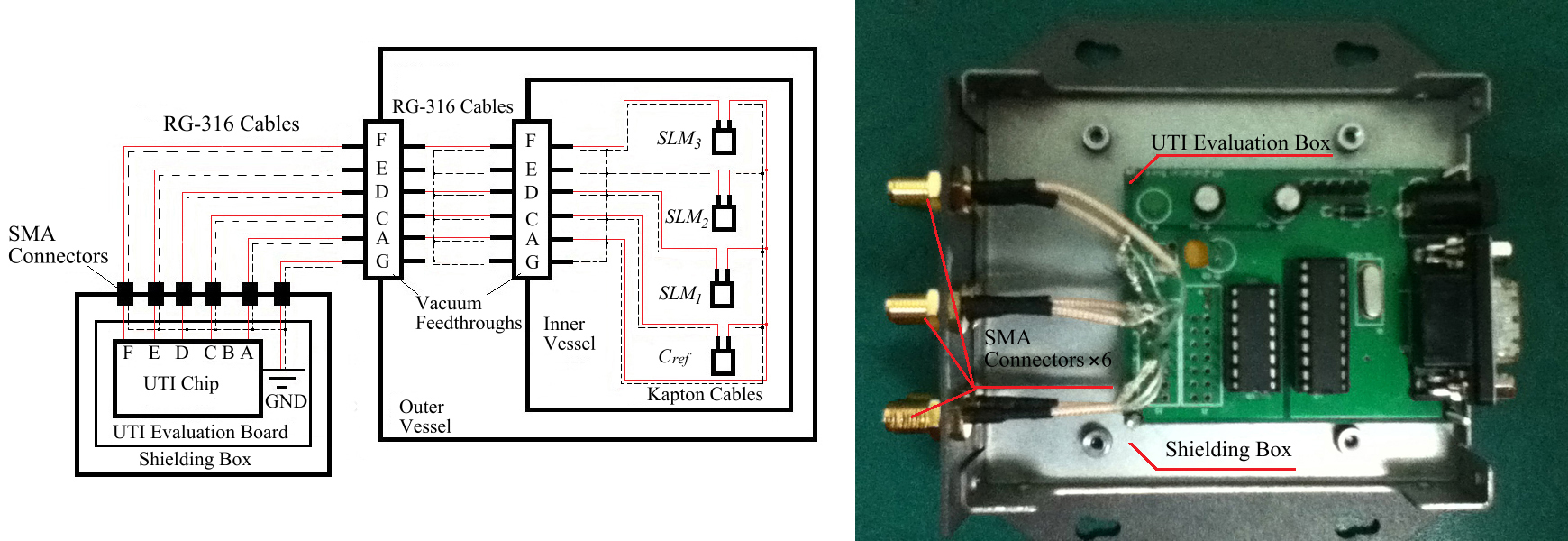}
\figcaption{\label{fig3}Readout schematics (left) and a picture (right) showing the cable connections for the three short level meters ($SLM_{1,2,3}$) and the reference capacitor ($C_{ref}$).}
\end{center}

The schematics to use mode 4 to measure the capacitance for the long level meter are shown in Fig.~\ref{fig4}. In this mode, a 300-pF reference capacitor is used. Unlike the reference capacitor for the short level meters, here we use a commercial capacitor mounted on the evaluation board outside of the vessel, instead of near the long level meter in the liquid xenon considering the radioactivity of the capacitor.

\begin{center}
\includegraphics[width=15cm]{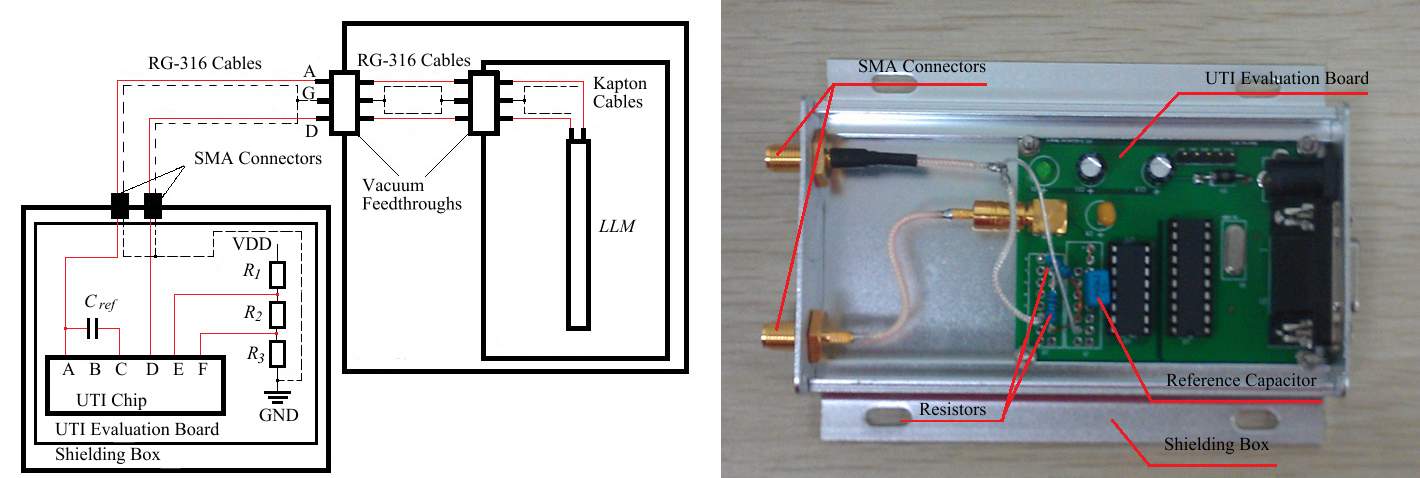}
\figcaption{\label{fig4}Readout schematics (left)  and a picture (right) showing the cable connections for the long level meter (LLM) and the reference capacitor ($C_{ref}$). The practical values for $R_1$ and $R_2$ are $25~k\Omega$ and $1~k\Omega$ respectively. $R_3$ is shorted in this readout mode.}
\end{center}

Coaxial cables are used to avoid capacitive-coupled interference. In Fig.~\ref{fig3} and Fig.~\ref{fig4}, red lines are signal wires of coaxial cables and the black dashed lines are their shielding layers. The shielding layers of the cables are connected together
to a common wire as ground. Although the cables go through two feedthroughs, their grounding shield layers are connected together. At last the ground wires are connected to the common ground on the UTI evaluation
board.

In the inner vessel, Kapton-insulated coaxial cables from the MDC Vacuum Products Company are used \cite{lab6},
which satisfy the high vacuum and cryogenic requirements. For the outer vessel which is used as a thermal insulation
chamber, the requirement of vacuum is not so strict and the surrounding temperature is close to the room temperature.
Therefore the RG-316 cables which are more common and cheaper are used for connections from the
inner vessel to the outer vessel. RG-316 cables are also used for the connections from the outer vessel to the readout
boxes at outside. The parasite capacitances from the long cables (about 5 meters for each level meter), feedthroughs and the plugs add in a total capacitance about several hundreds pF. By connecting the wires to the UTI board following the schematics in Fig.~\ref{fig3} and Fig.~\ref{fig4}, the parasite capacitances are mostly canceled.

\subsection{Readout software}
The capacitance measurement from the UTI chip can be read out, recorded and converted to the liquid level values by software with the UTI evaluation board connected to the computer by the RS-232 cables. Stand alone Windows-based software from Smartec Company can be used to read and record the measured values. We also developed Python-based software which is integrated to the PandaX slow control system to read and record the data. There are slow and fast modes to readout the capacitances. The during of one complete cycle of the output signal for the slow mode is about 100  ms, while the fast mode gives 10 ms duration. We choose the slow mode in the following measurement for a better resolution.

\section{Results and Discussion}

To check the accuracy and stability of the level meters, we performed several measurements in a small LXe
system before installing the level meters in the PandaX detector. The performances of the three short level meters
in a five-day operation are shown in Fig.~\ref{fig5} (left). The three short level meters first worked in vacuum for 47 hours. Then liquid xenon was filled into the detector and finally covered them completely. The short level meters were totally immersed in the liquid for 68 hours before the xenon was recuperated.

\begin{center}
\includegraphics[width=15cm]{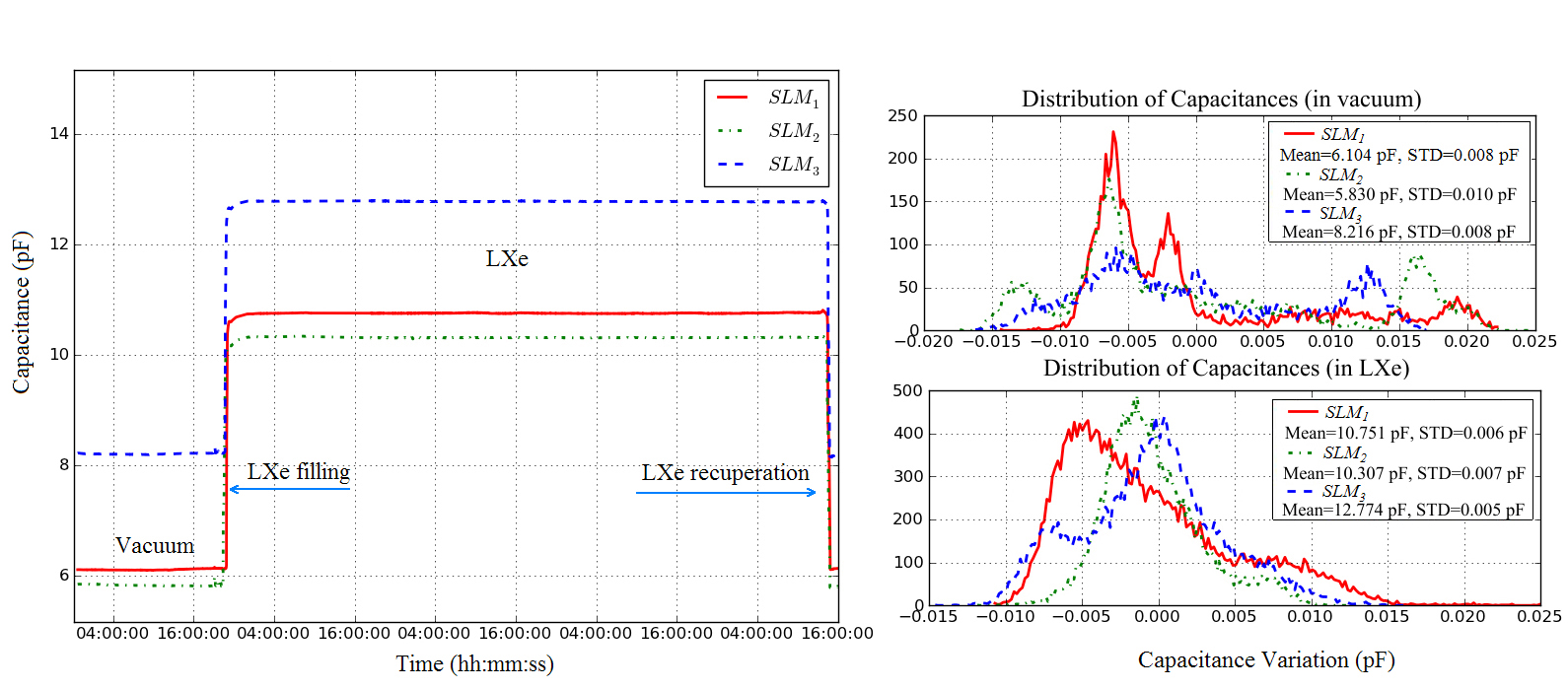}
\figcaption{\label{fig5} Left: Measured capacitances of the three short level meters ($SLM_{1,2,3}$)  during a five-day operation in a LXe system.  Right: Distributions of the measured capacitances, subtracting their mean values, for these level meters in vacuum and in liquid xenon. The mean and standard deviation (STD) values during the respective periods are listed in the figure legends.}
\end{center}

The mean values of the measured capacitances for the three level meters are listed in Table 2. The values are slightly higher than the expected theoretical values listed in Table 1. This is caused by the remaining parasite capacitances and the cumulative zero drift in the UTI board, especially large for the third level meter. However, it doesn't affect the measurement. The calibrated values, which are the capacitances per unit length, are calculated at 0.387, 0.371 and 0.380 pF/mm for the three level meters respectively, which are about 7-9\% lower than the expected theoretical values. To calculate the height of liquid level in the PandaX detector, the measured calibration values shall be used.

The measured fluctuations of capacitances represent the quality of the level meter structure and readout scheme. To quantify the stability and precision of the measurement, we use the standard deviation and peak to peak values during the entire period (68-hour) of measurement when the level meters are entirely immersed in liquid xenon. The measured values are shown in Fig.~\ref{fig5} (right) and listed in Table 2. The standard deviation values of the measured capacitances are between 0.005 to 0.007 pF, corresponding to a level fluctuation of less than 0.02 mm. The peak to peak values, represent the maximum fluctuation during the measured period, are 0.03 pF, corresponding to a maximum level readout fluctuation of less than 0.1 mm.
\newpage
\begin{center}
\tabcaption{ \label{tab2}Measured capacitances and their fluctuations of the three short level meters ($SLM_{1,2,3}$) and one long level meter ($LLM$). $C_{vac}$ and $C_{LXe}$ are the mean capacitance values when the level meters are in vacuum or fully immersed in liquid xenon respectively. The calibration value is the capacitance per unit length. $C_{STD}$ and $L_{STD}$ are the standard deviation values of capacitances and corresponding levels when the level meters are in liquid xenon during the measurement periods. $C_{P2P}$ and $L_{P2P}$ represent the largest fluctuations for the capacitances and corresponding levels during the measurement periods.}
\footnotesize
\begin{tabular*}{155mm}{@{\extracolsep{\fill}}cccccccc}
\toprule level meters &${C_{vac}}$(pF) & ${C_{LXe}}$(pF) & Calibration(pF/mm)  & ${C_{STD}}$(pF) & ${L_{STD}}$(mm) & ${C_{P2P}}$(pF) & ${L_{P2P}}$(mm)\\
\hline
$SLM_1$ &$6.104$ & $10.751$ &0.387 &0.006&0.016 &0.028& 0.09\\
$SLM_2$ &$5.830$ & $10.307$ &0.371 &0.007&0.019 &0.025& 0.07\\
$SLM_3$ &$8.216$ & $12.774$ &0.380 &0.005&0.013 &0.031& 0.08\\
$LLM$   &$120.0$ & $213.4$  &0.37  &0.05&0.14   &0.20& 0.53\\
\bottomrule
\end{tabular*}
\end{center}

\begin{center}
\includegraphics[width=15cm]{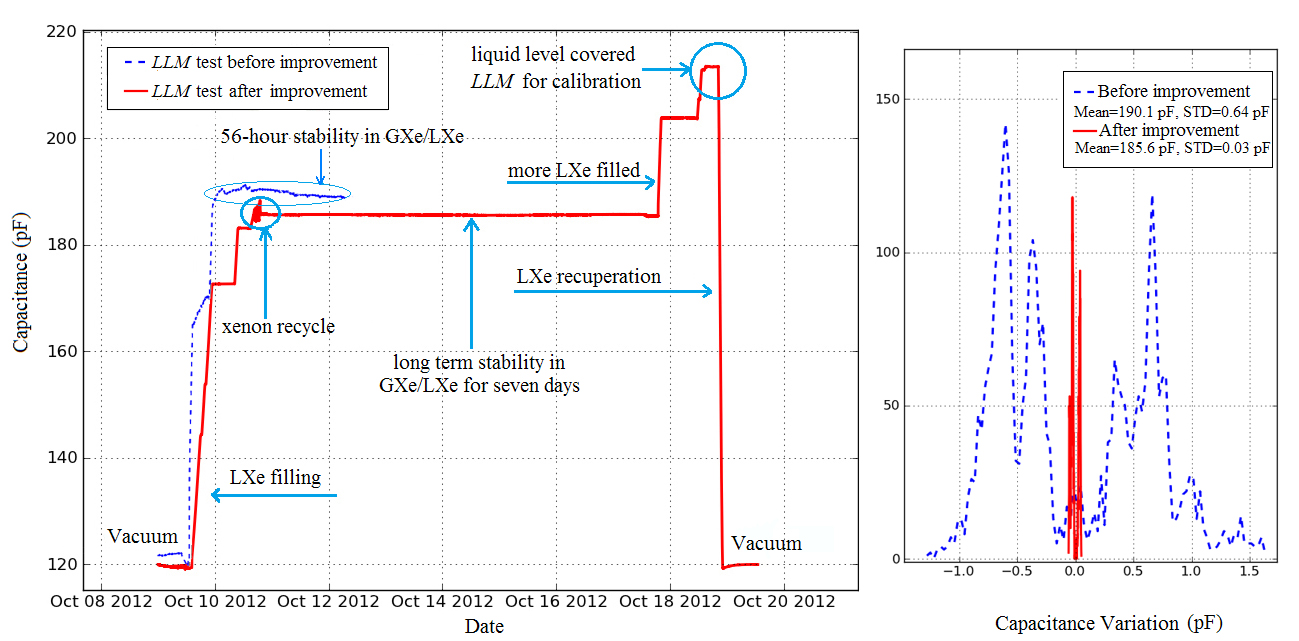}
\figcaption{\label{fig6} Left: Measured capacitances of the long level meter in a LXe system. A replacement of the reference capacitor significantly improved the stability of measurement. The blue dashed line shows the data before the improvement and the red line shows the data after the improvement.  Right: Distributions of the measured capacitances, after subtracting their mean values for the long level meter when the liquid xenon was in a stable condition. The reduction of the readout fluctuation is clear from the figure. The standard deviation value is reduced from 0.64 to 0.03 pF after the improvement.}
\end{center}

For the long level meter measurement, the first measurement showed a maximum level
fluctuation ($L_{P2P}$) up to several millimeters (see the blue dashed line in Fig.~\ref{fig6}), which was too large to satisfy the requirement of the experiment. After changing the electrical components on the readout board, we found that the quality of the reference capacitor affects dramatically of the precision of the measurement. The reference capacitor which
we use now is the 300 pF PFR series capacitor from Evox Rifa company. The resistors were replaced to the high precision ($\pm0.1\%~\Omega$)  0.25 W metal film resistors (RJ14)  and low-temperature-drift ($15\sim25~ppm/^{\circ}C$) type to further improve the measurement. The performance of the long level meter measurement after changing the reference capacitor and resistors is shown as the red line in Fig.~\ref{fig6}.

After the above mentioned improvement, the standard deviation value of the measured capacitance is 0.05 pF, corresponding to a 0.14 mm level fluctuation, when the long level meter was in a stable liquid xenon for seven days. The maximum fluctuation during the seven days is 0.2 pF (or 0.53 mm). After the seven days of operation, we filled more liquid xenon to cover the long level meter completely. Using the capacitance value measured when it's fully immersed in liquid xenon and in vacuum, we derived a calibration value of 0.37 pF/mm for the long level meter.

\section{Conclusion}

In this paper, we present a detailed design and development of the liquid level meters for the PandaX dark matter detector. Three short level meters to monitor the inclination of the TPC and one long level meter to monitor the overall liquid xenon level are developed. Special cares were taken for the cable connections and selection of the reference capacitors and resistors, which are critical for the stability and precision of the level measurement. The fluctuations (standard deviation) are less than 0.02 mm for the short level meters in a 68-hour test measurement, and is 0.14 mm for the long level meter in a seven-day measurement. The maximum fluctuations (peak to peak) during the corresponding periods are less than 0.1 mm for the short level meters and is 0.53 mm for the long level meter. The developed level meters and their readout method thus satisfy the requirement of the experiment and they have been installed in the PandaX detector.

\section*{Acknowledgment}
This work is supported within the PandaX experiment by the National Science Foundation of China (grant numbers: 11055003 and 11175117), the Science and Technology Commission of Shanghai Municipality (grant number: 11PJ1405300), and the Ministry of Science and Technology of China (grant number: 2010CB833005).

We would like to thank Dr. Angel Manzur for his valuable inputs and other members of the PandaX collaboration for useful discussions during the work.

\vspace{-1mm}
\centerline{\rule{80mm}{0.1pt}}

\clearpage

\end{CJK*}
\end{document}